Long Period Tidal Force Variations and Regularities in Orbital Motion of the

Earth-Moon Binary Planet System

Yu.N. Avsyuk

Institute Physics of the Earth RAS, Moscow, Russia

avsyuk@ifz.ru

L. A. Maslov

Aims College, Greeley, CO, USA

lev.maslov@aims.edu

and

Computing Center RAS, Khabarovsk, Russia

ms_leo@hotmail.com

**Abstract**

We have studied long period, 206 and 412 day, variations in tidal sea level corresponding to various moon phases collected from five observatories in the Northern and Southern hemispheres. Variations in sea level in the Bay of Fundy, on the eastern Canadian seaboard, with periods of variation 206 days, and 412 days, have been discovered and carefully studied by C. Desplanque and D. J. Mossman (2001, 2004). The current manuscript focuses on analyzing a larger volume of observational sea level tide data as well as on rigorous mathematical analysis of tidal force variations in the Sun-Earth-Moon system. We have developed a twofold model, both conceptual and mathematical, of astronomical cycles in the Sun-Earth-Moon system to explain the observed periodicity. Based on an analytical solution of the tidal force variation in the Sun-Earth-Moon system, it is shown that the tidal force can be decomposed into two components: the Keplerian component and the Perturbed component. The Perturbed component of the tidal force variation was calculated, and it was shown that the observed periodicity, 206 and 412 days, of atmospheric and hydrosphere tides results from variations of the Perturbed component of tidal force. The amplitude of the Perturbed component of tidal force is $19 \cdot 10^{-8} \, N/kg$. It is the same order of magnitude as the amplitude of the Keplerian component of tidal force: $58 \cdot 10^{-8} \, N/kg$. It follows that the Perturbed component of the variation of a tidal force must always be taken into consideration along with the Keplerian component in geodynamical constructions involving tides.

**Introduction**

Variation of the tidal force in the Sun-Earth-Moon system causes of a number of tidal cycles in the solid Earth, atmosphere, and hydrosphere. The following are the main cycles: principal lunar semidiurnal $M_2$, principal solar semidiurnal $S_2$, larger lunar elliptic semidiurnal $N_2$, lunar diurnal $K_1$, and lunar diurnal $O_1$. The longer period cycles are: lunar monthly $M_m$, lunisolar synodic fortnightly $M_{sf}$, and lunisolar fortnightly $M_f$. The lunar monthly $M_m$ cycle has duration 27.55 days, the lunisolar synodic fortnightly cycle has duration 14.76 days, and the lunisolar fortnightly cycle has duration 13.36 days.

The tidal cycles mentioned above can act as trigger mechanisms in the seismicity of the Earth and Moon. For example, study of the seismicity of the Earth under the influence of lunar tides (Tamrazian 1967) revealed that the maximum number of earthquakes happen when the Moon is at perigee and the minimum number of earthquakes occur when the Moon is at apogee. A thorough study of seismicity triggered by diurnal and semidiurnal tides has been conducted (Métivier et al. 2009). It was shown here that there exists a clear correlation, with 99% confidence, between the phase of the Moon and the timing of seismic events. The study of space regularities in the seismicity of the Earth and Moon (Levin, Sasorova 2010) revealed a similarity in the distribution of quakes in the Earth and the Moon. The quakes are distributed within two latitudinal belts: approximately 20° - 40°, on each side of the equator. Past research (Kosygin, Maslov 1986) discussed possible geodynamic consequences of solid Earth tides such as seismicity, drift of the lithosphere, and magnetic field generation.

Variations in the tidal sea level, with periods of variation 206 days, and 412 days, have been discovered and carefully studied by C. Desplanque and D. J. Mossman (2001, 2004). These authors studied variations of sea tides in the Bay of Fundy, on the eastern Canadian seaboard. However, diurnal, semidiurnal, and lunar monthly influences are more commonly discussed in the literature. The study of the seismicity of the Moon, conducted by the Apollo mission, showed that the seismicity of the Moon has a period of 27 days and also a period of 206 days (Lammlein 1974). Yu. Barkin and coauthors (Barkin et al. 2005) reported that the seismicity of the Moon has periods between 200 - 240 days and 360 - 440 days. Study of atmospheric air temperature variations in the Moscow region, (absolute values relative to the average) for the years 1960 - 2003, revealed periods of 27, 87, 206 and 355 days (Sidorenkov, 2009).

In this work we aim to:

a) Represent variations in tidal sea level corresponding to various moon phases collected from five different observatories worldwide in order to study the periodicity of these tides.

b) Develop a model of astronomical cycles in the Sun-Earth-Moon system to explain the observed periodicity in tidal sea level variations.

c) Apply an analytical solution for the tidal force variation that occurs in the Sun-Earth-Moon system to show that the tidal force can be decomposed into two components: the Keplerian component and the Perturbed component.

d) Calculate the Perturbed component and show that the observed periodicity, 206 and 412 days, of atmospheric and hydrosphere tides results from variations of the Perturbed component of the tidal force.

**Sea Tides and Their Interpretation**

Figure 1 shows graphs of syzygial sea tides in Murmansk, Russia; Puerto Williams, Chili; Suva, Fiji; Lerwick, Scotland; and Magadan, Russia from the years 1977 to 1979.

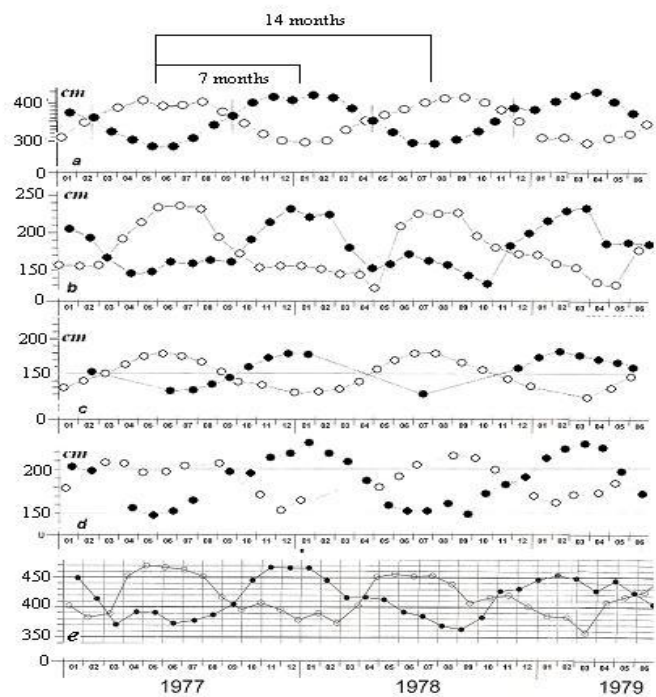

Figure 1
Heights of sygygial sea tides at full and new moons.
a - in Murmansk (Russia), 68.97 N, 33.08 E; b - in Puerto Williams (Chili), 54.93 S, 67.62 W;
c - in Suva (Fiji), 18.13 S, 178.43 E; d - in Lerwick (Scotland), 60.15 N, 1.15 W;
e - in Magadan (Russia), 59.57 N, 150.8 E; ○, ● - syzygial full moon and new moon tides.

As can be seen from these graphs, tides happen synchronously and independently at different observatories in the Northern and Southern hemispheres. It can also be seen that there exists a clear periodicity in tidal variations. The time interval between two closest full moon and new moon maximums is equal to 7 months (≈206 days), and the period of each of these variations is equal to 14 months (≈ 412 days). Figure 2 shows syzygial and quadrature sea level tidal variations for the years 1977 – 1983 in Murmansk, and the years 1977 – 1994 in Magadan. Here, as in Figure 1, syzygial and quadrature sea level tidal variations with periods 206 and 412 days can be clearly seen throughout the record. The quadrature tides are shifted relative to syzygial tides by 3.5 months (≈103 days).

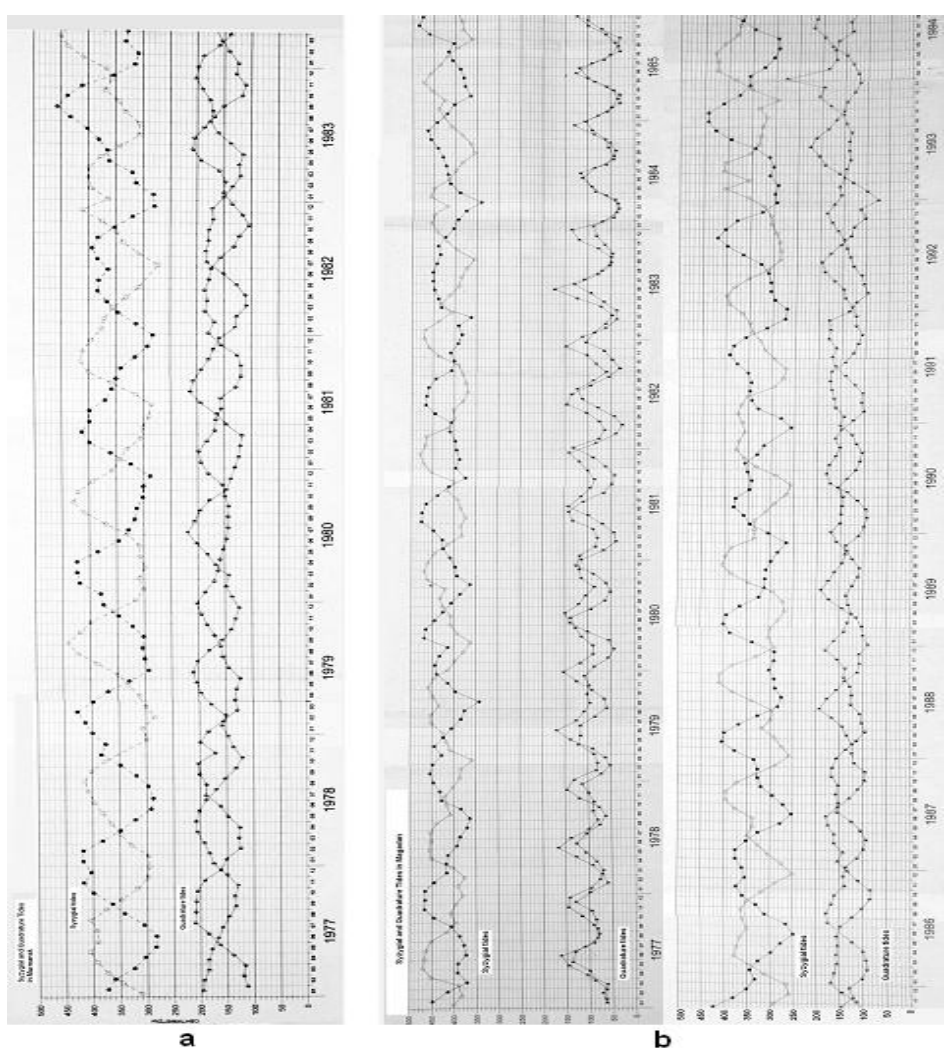

Figure 2
Heights of syzygial and quadrature tides in Murmansk (a), and in Magadan (b) for the time intervals 1977- 1983 and 1977 – 1994, correspondingly.

The data presented above have been kindly given to us by the following institutions: N.N. Zubov State Oceanographic Institute, www.oceanography.ru (Murmansk observatory); British Oceanographic Data Centre, UK, www.bodc.ac.uk, (Puerto Williams, Suva, and Lerwick observatories); Institute of Marine Geology and Geophysics FEBRAS, www.imgg.ru/rus/index.php, (Magadan observatory). To the data above can be added also long-period tidal sea level variations in the Bay of Fundy, eastern Canadian seaboard, from (Desplanque, Mossman 2001), figures 7 and 8.

The difference

$$H(t)_{fm} - H(t)_{nm} = H(t) \qquad (1)$$

(tide at full moon minus tide at new moon) calculated from observed data, Figure 1a, was approximated by the function:

$$D(t) = 120 \cdot sin\left(\frac{2\pi \cdot t}{14}\right)$$

with period T ≈ 14 months, or ≈ 412 days, Figure 3.

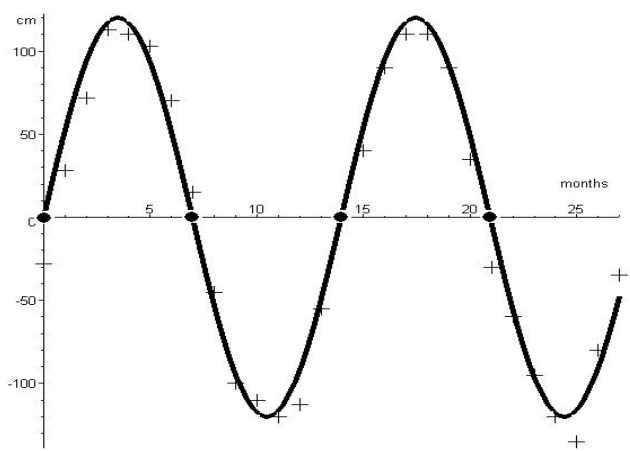

Figure 3
Difference in heights for full and new moon tides at Murmansk, Fig. 1a, formula (1).

A similar approximation was made for the other data in Figure 1. In all cases the best fit was shown to be for the period T ≈ 14 months.

To show more clearly the correspondence between the tidal and astronomical periodicities, zeros of the tidal function $H(t)$ and the mininal anomalistic months were plotted on the same graph, Figure 4. As seen from this Figure, the zeros of the function $H(t)$ correspond well to minimal anomalistic months of the motion of the Moon, and the Earth, with period 7 months (≈ *206*) days.

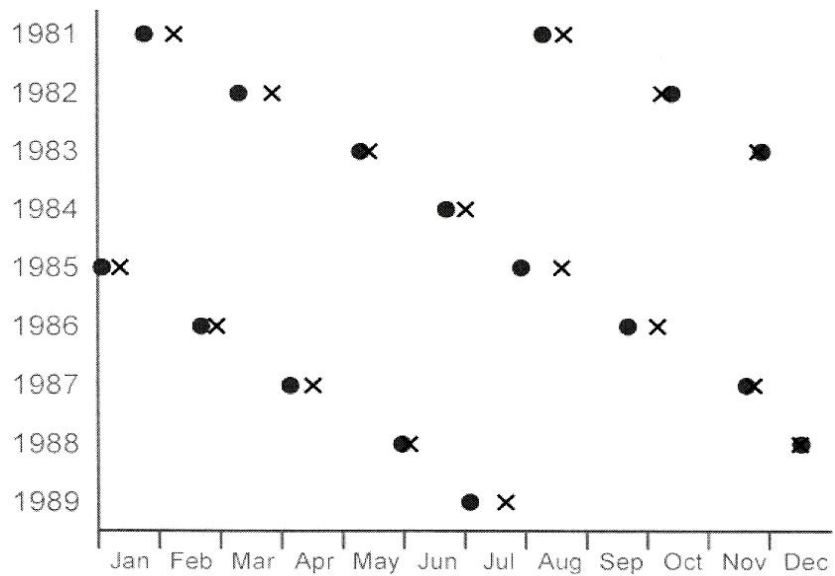

Figure 4
Zeros of the function $H(t)$ and of minimal anomalistic month distribution.
● - zeroes of the function $H(t)$, ✖ - minimal anomalistic month.

It can be seen from all the material presented, that the periodicity of the Earth's long period hydrosphere tides is the same as the periodicity of the variations of atmospheric air temperature, as well as the periodicity of lunar seismicity, 206 days as registered by the seismic stations of the Apollo mission.

**Cycles in Orbital Motion of the Earth and Moon Binary Planet System**

Study of the orbital motion of the Earth-Moon barycenter around the Sun, and the orbital motion of the Earth and Moon centers around their common barycenter, (Avsyuk 1996; Avsyuk and Suvorova 2006) showed that the time from one full moon in perigee until the next full moon in perigee is the same as the time from one high tide to another, and is 412 days. Let $T_1$ be the time that we observe a full Moon with the entire system in perigee, as shown in Figure 5.

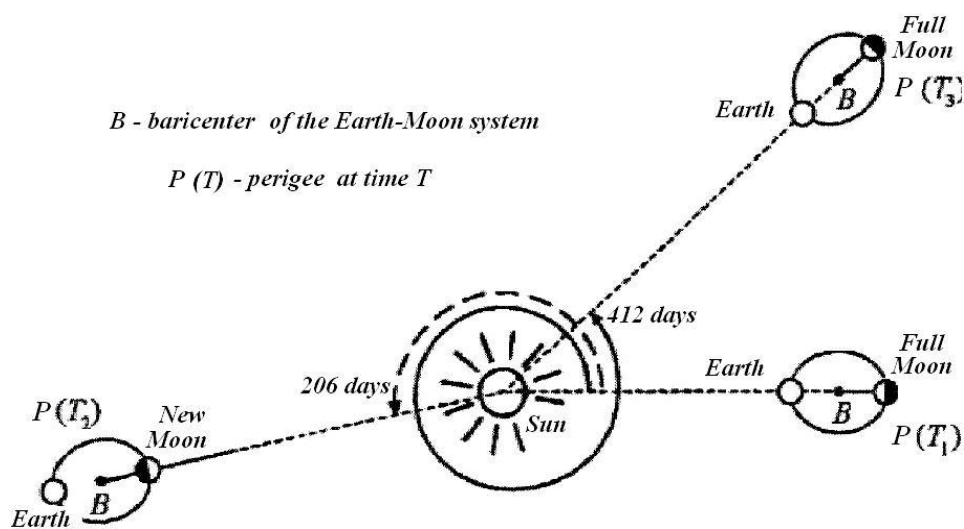

Figure 5
Illustration of the time recurrence of full moon coinciding with the passage through the orbital perigee. Since the major axis of the Moon's (Earth's) orbit is not fixed in space (the period of perigee revolution is 3230.25 days), the full moon phase and the passage through perigee of the full moon coincide at the time moment $T_3 = T_1 + 412$ days. The time moment $T_2$ fixes the new moon that coincides with the passage through the perigee).

In 206 days, one half of the total 412 day period, we will again have orbital perigee of the system, but with a new moon phase instead of a full moon phase; let us call this time $T_2$. The original configuration will occur at the time $T_3$, 412 days after $T_1$, Figure 5. This is caused by the fact that the Earth-Moon orbit is not fixed in space and the period of its perigee revolution is 3230.25 days. This reasoning can help us to understand the nature of the 206 and 412 day tidal cycles on a qualitative level.

**Variation of a Tidal Force in the Earth-Moon-Sun System. Mathematical Solution.**

The motion of the Earth-Moon binary planet system around the Sun can be represented as the sum of two components: the orbital motion of the Earth-Moon barycenter around the Sun (Keplerian motion), and the motion of the Earth and Moon around their common barycenter (Perturbed motion). The Perturbed motion describes the periodic motion of the center of mass of the Earth in its elliptical orbit around the Earth-Moon barycenter. In astronomy, this phenomena is well known as the lunar inequality, and has angular variation 6.44''.

The Perturbed motion of the Earth in the gravity field of the Moon and the Sun produces an additional component of tidal force, which we call the Perturbed component. In this approach, the total tidal force, acting on the Earth, is a sum of two components: Keplerian and Perturbed. The theory of the tidal force variations in cases of unperturbed (Keplerian) and Perturbed motion of a body was first developed in the work by Yu. N. Avsyuk (Avsyuk, 1996, 2001). The total tidal force as a function of the positions of the Moon, Sun, and Earth relative to each other was obtained as a solution of the system of equations of motion in Lagrange form. The tidal force was represented as a sum of Keplerian, $\bar{K}$, and Perturbed, $\bar{P}$, components:

$$\bar{F} = \bar{K} + \bar{P} \qquad (2)$$

The Keplerian component is described by the formula:

$$\bar{K} = G \cdot \delta m \cdot \left( \frac{M_2}{\rho_1^3} \cdot \bar{\rho}_1 - \frac{M_2}{R_1^3} \cdot \bar{R}_1 \right) + G \cdot \delta m \cdot \left( \frac{M_3}{\rho_2^3} \cdot \bar{\rho}_2 - \frac{M_3}{R_0^3} \cdot \bar{R}_0 \right) \qquad (3)$$

The Perturbed component is written in the form:

$$\bar{P} = C \cdot \left( \frac{3}{2} \cos(2D + 0.5) \cdot \bar{r} + \sin(2D) \cdot \bar{\tau} \right) \qquad (4)$$

$$C = G\delta m \frac{R_1 M_2 M_3}{(M_1 + M_2) \cdot R_0^3}$$

$M_1, M_2, M_3$ are masses of the Earth, Moon and Sun, $\rho_1$ is the distance from the center of the Moon to an arbitrary point in the Earth, $\rho_2$ is the distance from the center of the Sun to an arbitrary point in the Earth, $R_0$ is the distance from the Earth-Moon barycenter to the Sun, $R_1$ is the distance between the gravity centers of the Earth and Moon. Terms of order $(R_1 / R_0)^2$ are omitted in the derivation of (3) and (4). In this formula $\bar{r}$ and $\bar{\tau}$ are unit vectors in the radial and tangential directions respectively, $D = \lambda_S - \lambda_M$, $\lambda_S$ and $\lambda_M$ are longitudes of the Sun and Moon. A computer program was written to study the variation of tidal force for different input parameters. A graph of the function $|\bar{P}|/C$, calculated with two week intervals, is shown in Figure 6. The long period variations of a tidal force with periods ≈ 206 and ≈ 412 days are well displayed by the envelope function clearly visible in Figure 6.

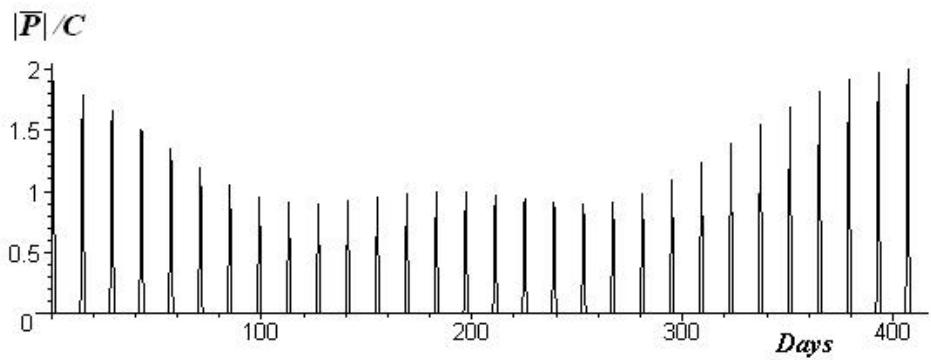

Figure 6
Graph of the Perturbed component of tidal force $|\overline{P}|/C$, calculated with two week intervals.

It is worthwhile to note, that the amplitude of the 206 day cycle is approximately 1.803 times less than the amplitude of the 412 day cycle. The amplitude of the Perturbed component of tidal force is $19 \cdot 10^{-8} \, N/kg$. It is the same order of magnitude as the amplitude of the Keplerian component of tidal force: $58 \cdot 10^{-8} \, N/kg$. It follows that the Perturbed component of tidal force variation must always be taken into consideration along with the Keplerian component in geodynamical constructions involving tides.

The formula (4) does not contain angular distances between heavenly bodies and a celestial equator. This means that the declination of the Moon and Sun relative to the Earth's equator is a secondary factor for the generation of tidal periodicities of 206 and 412 days. This fact was mentioned also by C. Desplanque and D.J.Mossman, (2004), figure 68, p.106 in their study of the Bay of Fundy tides.

**Conclusion**

We have presented a large amount of new data: syzygial and quadrature tidal sea level variations for the years 1977 – 1985 registered at five different observatories (Murmansk, Magadan, Lerwick, Puerto Williams, and Suva) situated in the Nothern and Southern hemispheres. We have shown that these tides have periods of 206 and 412 days; they are independent of the geographical position of the observatory and synchronous in time. To study this phenomena, we developed a mathematical theory of the tidal force variation in the Earth-Moon-Sun system and from the solution evaluated periods of the Perturbed tidal force variations and its magnitude. We have shown from this solution that the long period variations of the Perturbed tidal force component have periods of ≈ 206 and ≈ 412 days, which are in the excellent correlation with periods and regularities in the orbital motion of the Earth and Moon as a binary planet system. We have also shown from this solution that the magnitude of the Perturbed component of tidal force is $19 \cdot 10^{-8} \, N/kg$. It is very important that this force is only three times smaller than the magnitude of the Keplerian component of the tidal force $58 \cdot 10^{-8} \, N/kg$. It follows that the Perturbed component of tidal force variation must always be taken into consideration along with the Keplerian component in geodynamical constructions involving tides and the other planetary processes.

We have discussed the results of other authors, that show the long period variations of a tidal force have a significant impact not only on the Earth's hydrosphere, but also on the Earth's atmosphere and the Earth and Moon interiors. The long period variations of the Earth and Moon seismicity, with periods of ≈ 206 and ≈ 412 days, are the direct manifestation of corresponding variations of the tidal force in this system.

This work was supported by RFBR grant 09-05-00426-a.

**Acknowledgments**

We are grateful to the Editor and reviewers of this article for their remarks, comments and patience, contributing to the considerable improvement of our work.